\documentclass[]{spie}  

\usepackage{lineno}
\usepackage{amsmath,amsfonts,amssymb}
\usepackage{graphicx}
\usepackage{subcaption}
\usepackage[colorlinks=true, allcolors=blue]{hyperref}

\title{The Simons Observatory: Overview of the Cryogenic Half-wave Plate Polarization Modulators}

\author[a]{Junna Sugiyama}
\author[b]{Kyohei Yamada}
\author[c]{Bryce Bixler}
\author[d]{Daichi Sasaki}
\author[e]{Yuki Sakurai}
\author[c, f]{Kam Arnold}
\author[g, h]{Nicholas Galitzki}
\author[d, i, j, k]{Akito Kusaka}

\affil[a]{Department of Physics and Astronomy, University of California Riverside, Riverside, CA 92521, USA}
\affil[b]{Joseph Henry Laboratories of Physics, Jadwin Hall, Princeton University, Princeton, NJ 08544, USA}
\affil[c]{Department of Physics, University of California San Diego, La Jolla, CA 92093, USA}
\affil[d]{Department of Physics, The University of Tokyo, Tokyo 113-0033, Japan}
\affil[e]{Department of Mechanical and Electrical Engineering, The Suwa University of Science, Nagano 391-0213, Japan}
\affil[f]{Department of Astronomy and Physics, University of California San Diego, La Jolla, CA 92093, USA}
\affil[g]{Department of Physics, University of Texas at Austin, Austin, TX, 78712, USA}
\affil[h]{Weinberg Institute for Theoretical Physics, Texas Center for Cosmology and Astroparticle Physics, Austin, TX 78712, USA}
\affil[i]{Physics Division, Lawrence Berkeley National Laboratory, Berkeley, CA 94720, USA}
\affil[j]{Research Center for the Early Universe, School of Science, The University of Tokyo, Tokyo 113-0033, Japan}
\affil[k]{Kavli Institute for the Physics and Mathematics of the Universe (WPI), UTIAS, The University of Tokyo, Kashiwa, Chiba, 277-8583, Japan}

\authorinfo{Send correspondence to Junna Sugiyama. E-mail: junna.sugiyama@gmail.com}

\begin{document} 
\maketitle
\begin{abstract}

The Simons Observatory (SO) is a ground-based Cosmic Microwave Background (CMB) experiment that is located in the Atacama plateau.
The Small Aperture Telescopes (SATs) of SO are optimized for polarimetry on the degree scale. 
Atmospheric $1/f$ contamination of the CMB signal poses a significant challenge for observations at this angular scale.
In order to control the $1/f$ noise, the SATs utilize a Cryogenic Half-Wave Plate (CHWP) in their optics.
The CHWP modulates the polarization signal to a higher frequency to separate it from the unpolarized atmospheric noise.
Precision measurements of the CHWP rotation angle are required to successfully recover the target polarized signals.
We present a method to reconstruct the CHWP rotation angle that achieves a noise level of 0.16\,$\mu\mathrm{rad\sqrt{s}}$, meeting our requirement.

\end{abstract}


\section{INTRODUCTION}
\label{sec:intro}  
The Cosmic Microwave Background (CMB) is the oldest observable electromagnetic radiation, which was released at the epoch of recombination, approximately 380,000 years after the birth of the universe.
Scalar perturbations at the recombination era generated primary anisotropies of CMB polarization, with a possible additional contribution from tensor perturbations.
While scalar perturbations of the early universe have been well characterized by previous CMB observations, tensor perturbations have not been detected yet.
Primordial tensor perturbations produce a degree-scale parity-odd anisotropy pattern in the CMB polarization, which is called a $B$-mode.
The amplitude of primordial $B$-mode polarization is parametrized by the tensor-scalar ratio\cite{1967ApJ} $r$; the current upper limit of $r$ is 0.036 \cite{latest_r}.

The Simons Observatory (SO) is the largest ground-based CMB experiment in history.
SO has been observing since October 2023.
SO aims to map the sky at millimeter wavelengths with unprecedented sensitivity to explore $B$-mode polarization corresponding to $\sigma (r)=0.003$ with five-year survey, and $\sigma (r)=0.001$ with ten-year survey~\cite{Ade_2019, SOforecast2}.
SO is located on Cerro Toco in the Atacama Desert, Chile (22$^{\circ}$57$^{\prime}$S, 67$^{\circ}$47$^{\prime}$W, 5200\,m a.s.l.)~\cite{Ade_2019}.
The Small Aperture Telescopes (SATs) of SO are dedicated to the degree-scale $B$-mode search \cite{SAT1, SAT2}.
Three SATs are currently taking data: SAT1 and SAT3 employ dichroic detectors at 90/150 GHz, while SAT2 operates at 220/280 GHz. 

\begin{figure}
    \centering
    \begin{subfigure}{0.45\textwidth}
        \centering
        \includegraphics[width=\textwidth]{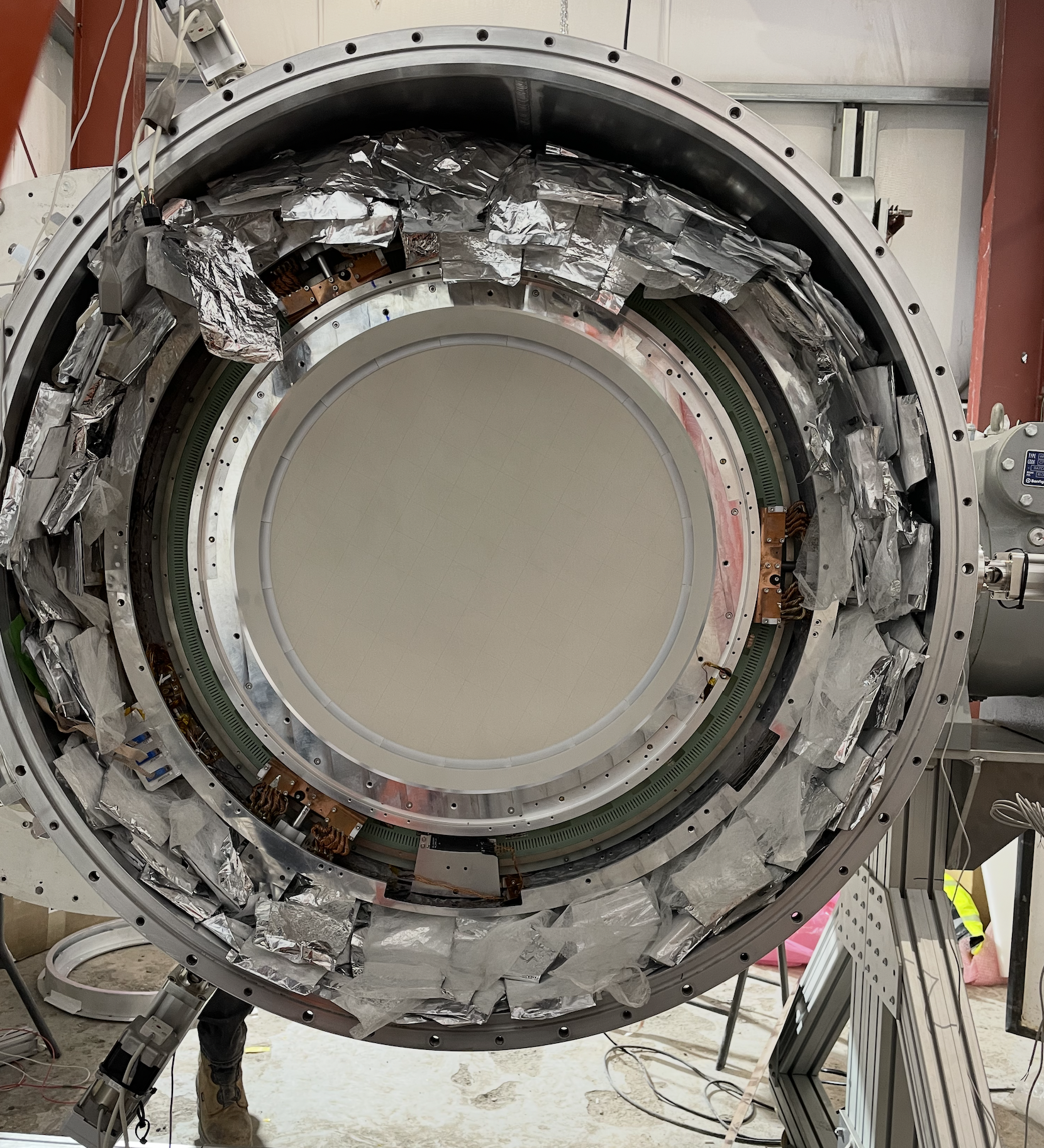}
        \caption{Top view of the CHWP installed in SAT2. 
        The slotted circular plate is the encoder plate and the trapezoid on the bottom is one of the encoder heads. 
        The configuration of the telescope rotates by 90\,deg during the observation.}
        \label{fig:chwp1}
    \end{subfigure}
    \hfill
    \begin{subfigure}{0.53\textwidth}
        \centering
        \includegraphics[width=\textwidth]{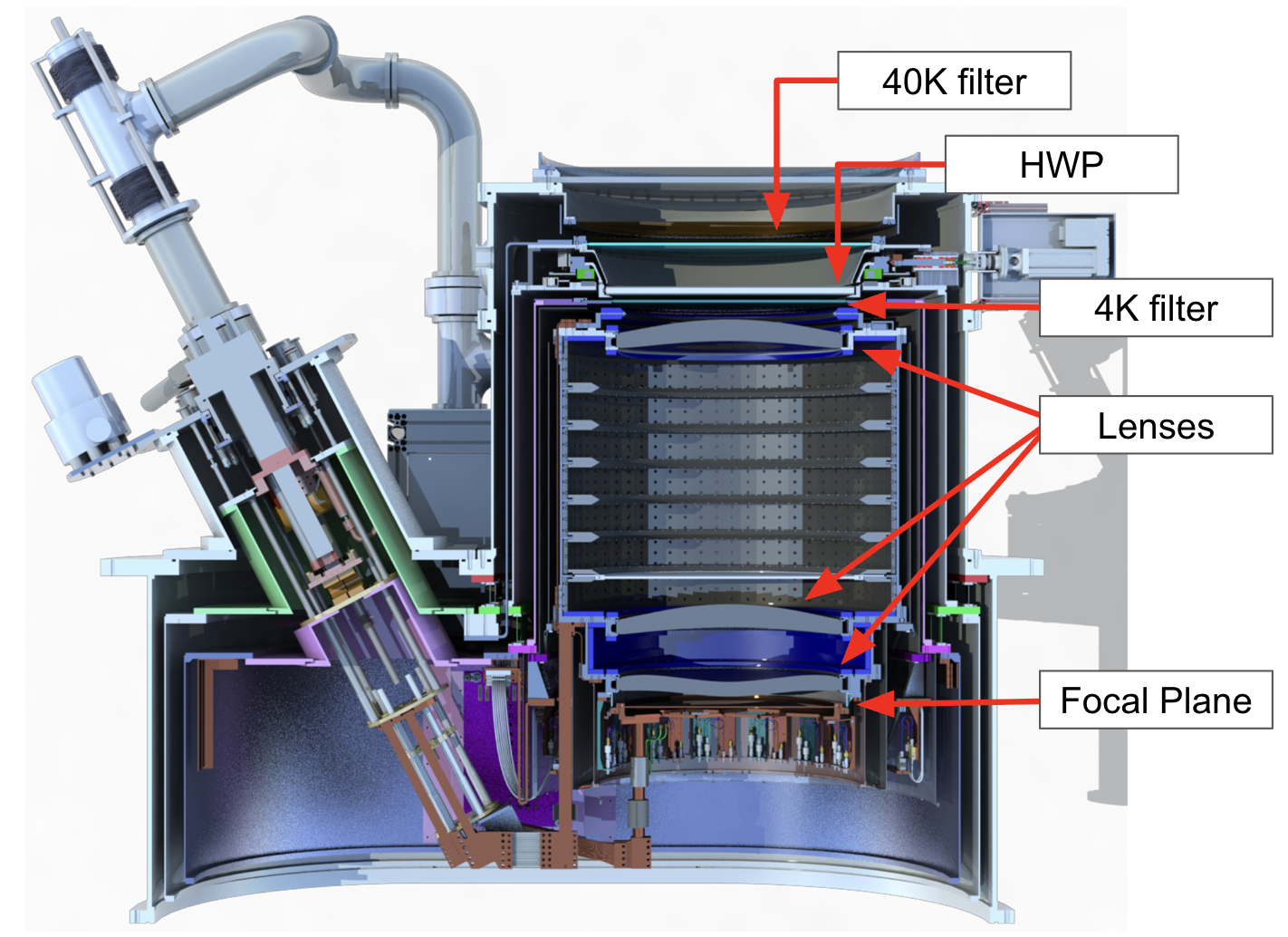}
        \caption{The cross section of the SAT receiver. The CHWP is located on the 40\,K stage, right after the 40\,K filter.\\
        ~\\~\\~}
        \label{fig:crosssection}
    \end{subfigure}

    \caption{The CHWP assembly installed in the SATs.}
    \label{fig:two_pngs}
\end{figure}

One of the largest challenges in the degree-scale $B$-mode search is the low-frequency noise, also known as $1/f$ noise.
The observed sky maps are reconstructed from the time-ordered data of the telescopes scanning the sky.
The low-frequency noise in the time-ordered data contaminates the degree-scale fidelity of the reconstructed map.
The SATs use a Cryogenic continuously rotating Half-Wave Plate (CHWP) as a polarization modulator (Figure~\ref{fig:chwp1}).
A HWP is a birefringent optical element designed to introduce a relative phase shift of $\pi$ between two orthogonal polarization components at a target wavelength.
A HWP rotating at frequency $f$ modulates the incident linear polarization at $4f$.
The signal modulated by the HWP, $d_m(t)$, is expressed in terms of the HWP crystal axis angle $\chi(t)$:
\begin{equation}\label{eq:modulation}
    d_m(t) = \mathcal{T}\cdot I(t) + \varepsilon \mathcal{T}\cdot\mathrm{Re}[(Q(t)+iU(t))m(\chi)]+N(t),
\end{equation}
where $t$ is time, $\mathcal{T}$ is the transmission of the HWP, $\varepsilon$ is the modulation efficiency of the HWP, $I(t)$/$Q(t)$/$U(t)$ are the Stokes parameters of the incident light, and $N(t)$ is the noise on the detector, respectively.
The modulation function $m(\chi)$ is expressed as
\begin{equation}
    m(\chi) = \exp(-4i\chi).
\end{equation}
To reconstruct the input $Q$ and $U$, we bandpass filter the modulated data around the modulation frequency, multiply it by the demodulation function $2m(-\chi)$, and then apply a low-pass filter to it. 
This procedure, which is called demodulation \cite{Kusaka_2014}, reproduces the input polarization signal as
\begin{equation}\label{eq:demodulation}
    d_d(t) = \varepsilon \mathcal{T}(Q(t)+iU(t))+N_d(t).
\end{equation}

\begin{figure}
    \centering
    \includegraphics[width = 0.5\textwidth]{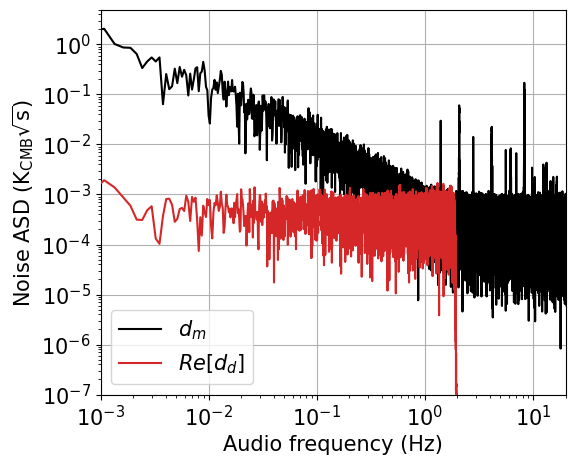}
    \caption{The noise amplitude spectra of a SAT 150\,GHz detector from an example 1 hour observation. The precipitable water vapor during this measurement was 0.6\,mm. The black line shows the noise before demodulation, $d_m$. The $1/f$ noise is dominant at low-frequencies, while the modulated polarization peaks at 8\,Hz. Besides the modulated polarization, there are peaks at $2n$\,Hz ($n=1, 2, ...$), which we call HWPSS (see Section~\ref{sec:HWPSS}).
    The red line represents the residual noise in the demoduled signal, $d_d$. The demodulated signal is cut off at 2\,Hz since the modulation innately down samples the signal to the HWP rotation frequency.}
    \label{fig:chwp2}
\end{figure}

Figure~\ref{fig:chwp2} shows the amplitude spectrum density (ASD) of a 150\,GHz detector data during the CMB observation.
The $d_m$ is represented by the black line.
The frequencies at $<$1\,Hz are mostly made up of $1/f$ noise.
This contaminates the measurement at angular scales $>$0.1\,deg, given the nominal SAT scan speed.
The SATs rotate the HWP at 2\,Hz. 
It modulates the polarization signal to 8\,Hz, above the $1/f$ noise.
The $d_m$ is bandpass filtered around 8$\pm$1.9\,Hz.
The demodulation returns $d_d$, from which the $Q$ and $U$ are obtained.
Thus the modulation and demodulation decompose the signal into the unpolarized noise and polarization. 
This technique also enables a single detector to measure both $Q$ and $U$ polarization simultaneously.
The residual polarized noise in $\mathrm{Re}[d_d]$ is represented by the red line in Figure~\ref{fig:chwp2}.
The noise amplitude around 0.1\,Hz is reduced by two orders of magnitude, and it is white noise dominated.

The instrumental polarization which arises toward the detector side of the CHWP is not modulated and thereby separated from the polarization from the sky.
For this reason, the CHWP is placed as skyward in the optical chain as possible, behind only the window and two infrared filters, as shown in Figure~\ref{fig:crosssection}.
This also simplifies the cryogenic thermal design compared to locating it on the colder stages, while keeping the thermal loading on the detectors acceptably low. 
As an added benefit, maintenance access is less demanding at this location than at colder stages.
The design of the cryogenic rotation mechanism is mostly common among the SATs, while the HWP optics are optimized depending on the spectral band covered by the telescopes\cite{yamada2023simons, Sugiyama2024}.
The CHWP for the 30/40\,GHz SAT, which will be deployed in 2027, has a redesigned rotation mechanism with enhanced bearing force~\cite{Sasaki_2025}.

\section{INSTRUMENT}
The CHWP of the SATs operates at 50\,K to mitigate the thermal loading on the detectors.
It has an optical diameter of 490\,mm, which to our knowledge is the largest for a continuously rotating CHWP in a CMB experiment.
In order to achieve the rotation of the large diameter HWP at a cryogenic temperature, the rotation mechanism employs the Superconducting Magnetic Bearing (SMB).
The pinning force of the SMB levitates the HWP and fixes its rotation axis.
The cryogenic temperature of the floating HWP is maintained by the radiation cooling.
The SMB enables smooth rotation with little friction, while keeping the large open aperture.
It can also avoid the stress from the differential thermal contraction, which is inevitable in a mechanically contacted rotation mechanism.

\begin{figure}
    \centering
    \begin{subfigure}{0.48\textwidth}
        \centering
        \includegraphics[width=\linewidth]{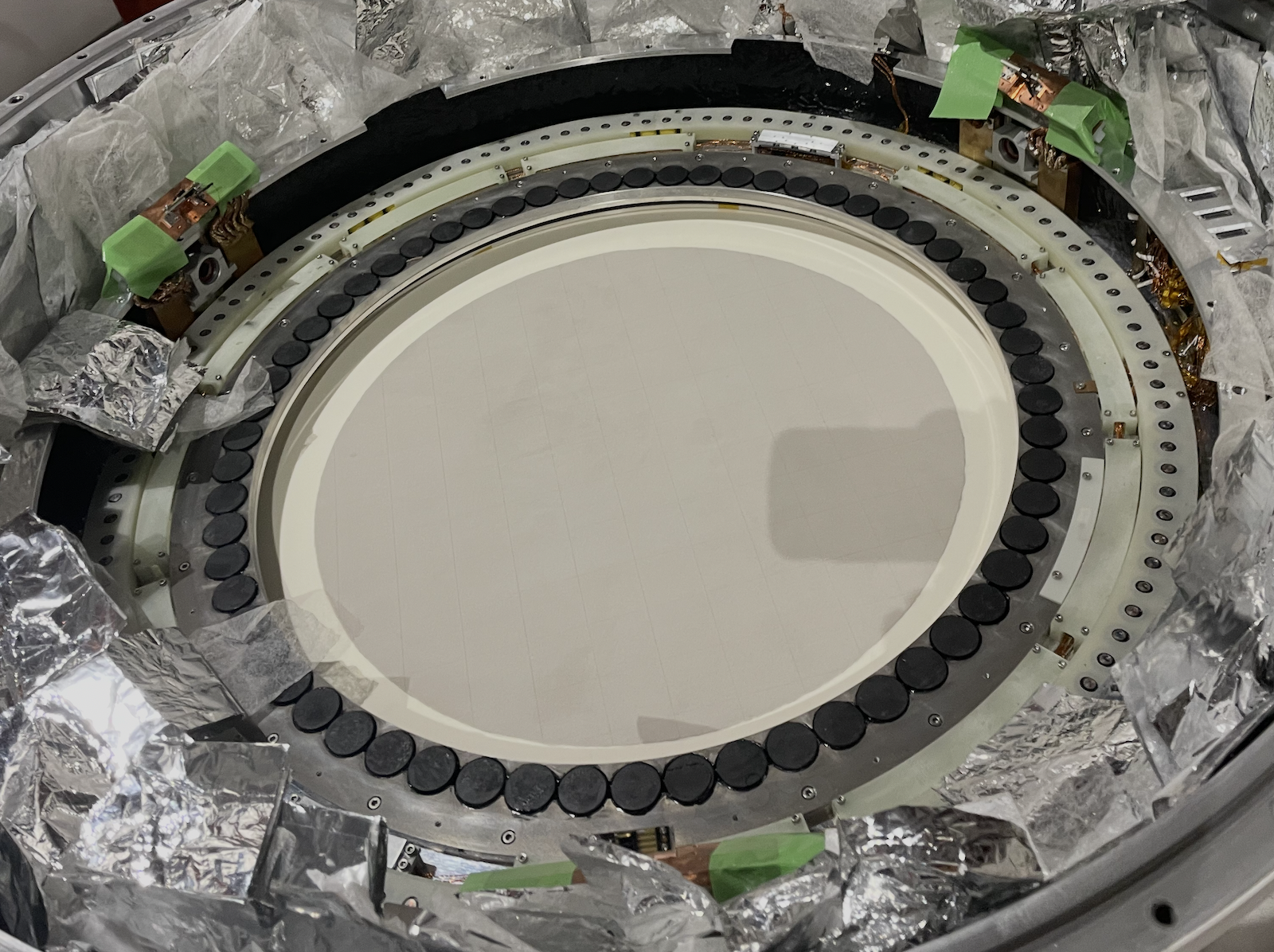}
        \caption{The stator of the CHWP. The black pucks are YBCO. The solenoids array is placed outside of the YBCO ring. The 4\,K filter is located below the stator.\\~\\~\\~}
        \label{fig:stator}
    \end{subfigure}
    \hfill
    \begin{subfigure}{0.44\textwidth}
        \centering
        \includegraphics[width=\linewidth]{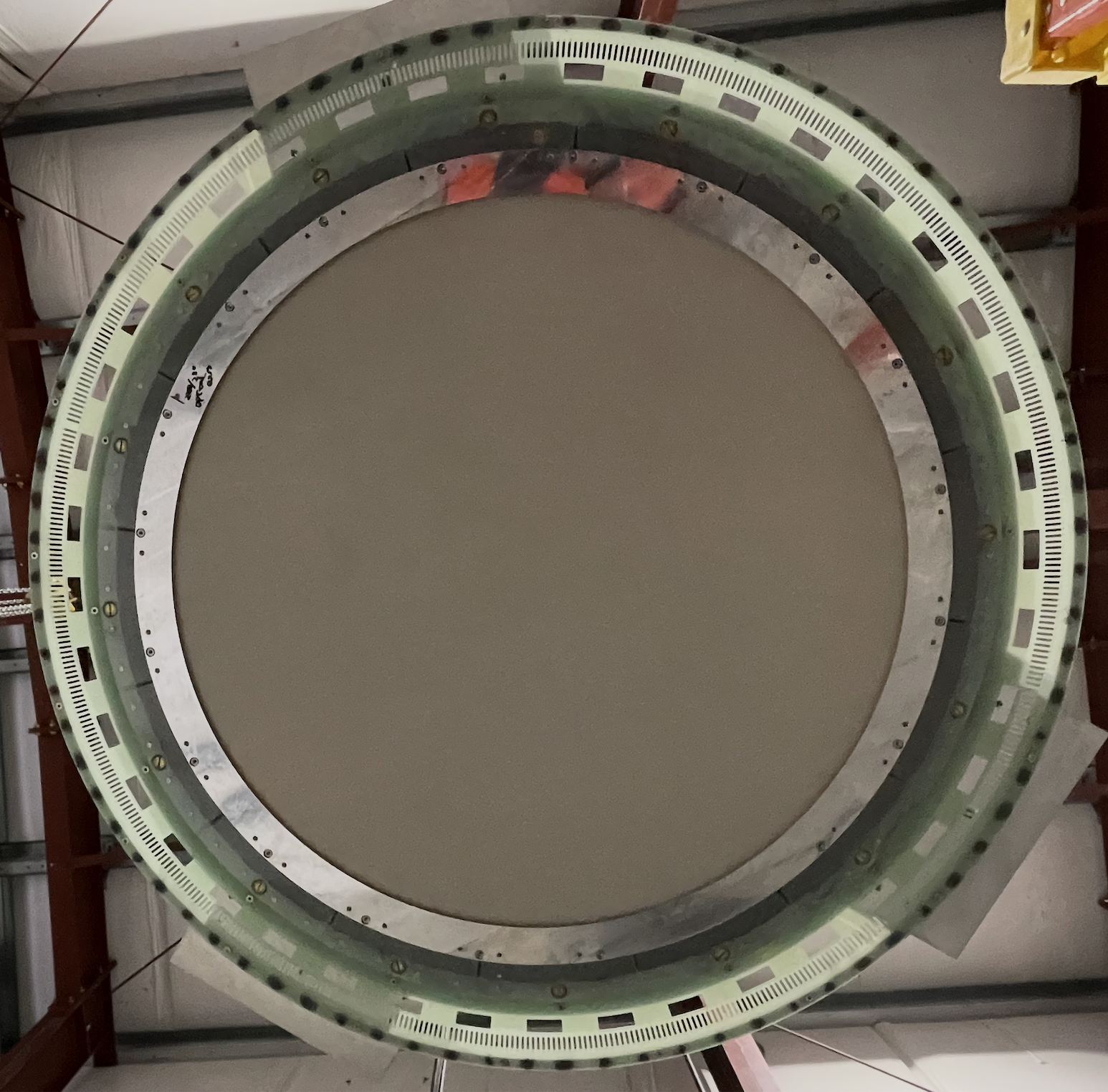}
        \caption{The bottom side of the rotor. The neodymium magnet surrounds the HWP optics. The encoder plate is attached to the rotor. There are sprocket neodymium magnets on the edge of the encoder plate, which generates the rotor torque by coupling to the solenoids on the stator.}
        \label{fig:rotor}
    \end{subfigure}
    \vspace{0.5cm}
    \begin{subfigure}{0.5\textwidth}
        \centering
        \includegraphics[width=\textwidth]{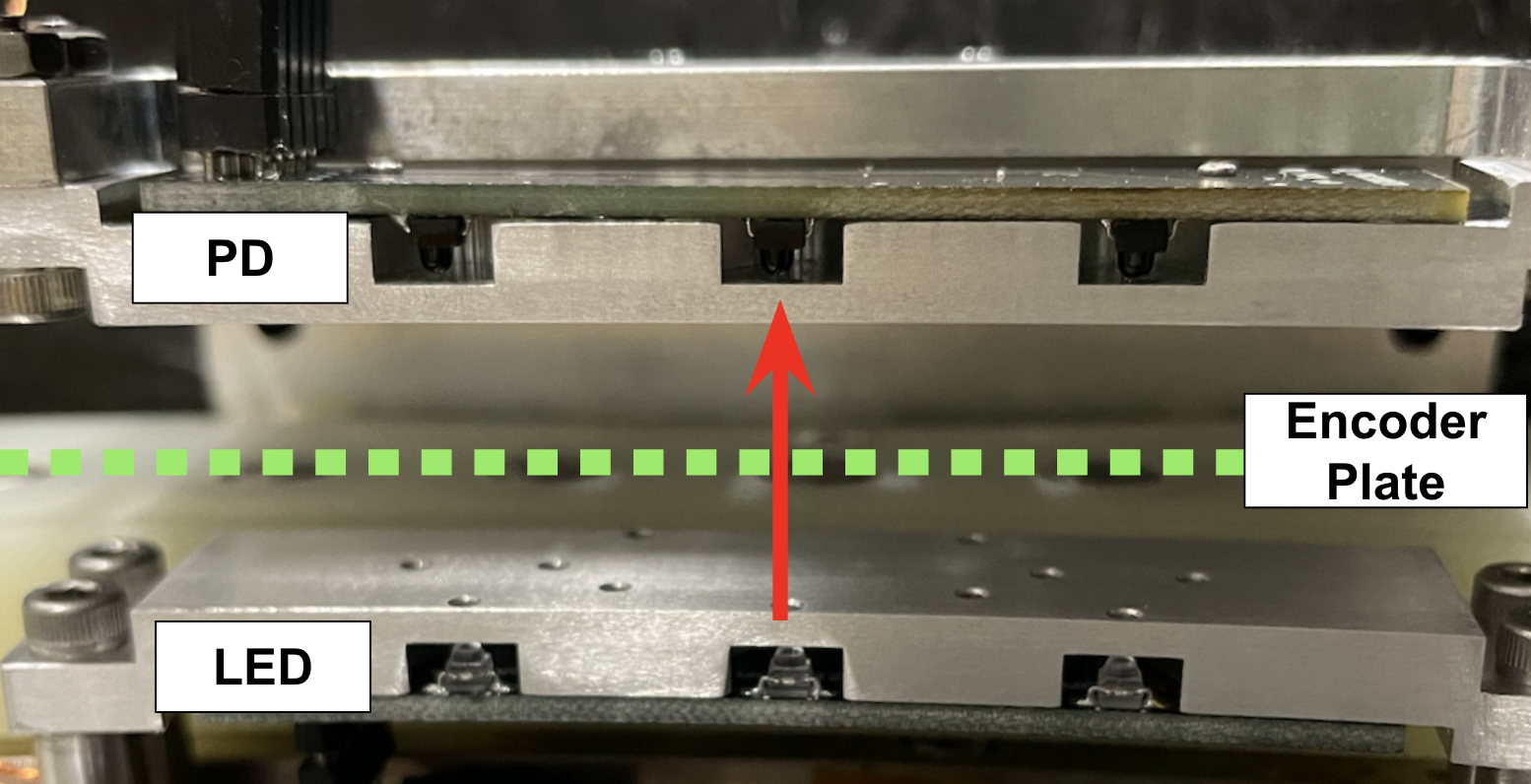}
        \caption{A magnified view of the optical encoder assembly. The IR signal emitted by the LED is chopped by the encoder plate. There are five PDs and five LEDs on one optical encoder assembly. Two PD/LED sets are dedicated to the angle reading, while the other three PD/LED sets are for the CHWP motor control. }
        \label{fig:encoder}
    \end{subfigure}
    \hfill
    \begin{subfigure}{0.4\textwidth}
        \centering
        \includegraphics[width=\textwidth]{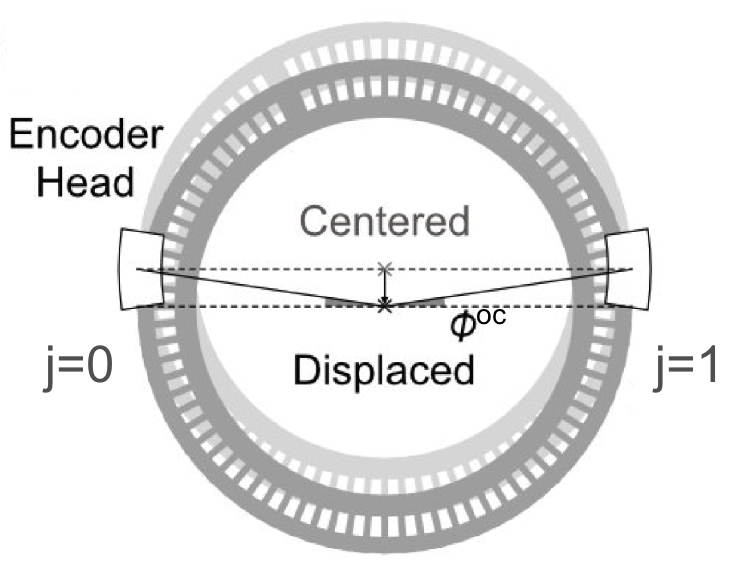}
        \caption{Schematic view of the optical encoder system, highlighting that when the HWP and the encoder plate are off-centered perpendicular to the line connecting the encoders. In such cases, the encoder incorrectly measures the rotation angle by $\phi^{oc}$. }
        \label{fig:sag_model}
    \end{subfigure}
    \caption{The rotation mechanism of the CHWP.}
    \label{fig:rotation_mechanism}
\end{figure}

\subsection{Rotation Mechanism}
The rotation mechanism is composed of rotor and stator sections.
They are held together through a 550\,cm diameter SMB. 
The stator has Yttrium Barium Copper Oxide (YBCO) pucks, a kind of type-II superconductor (Figure~\ref{fig:stator}), while the rotor has a ring-shaped neodymium magnet assembly on its bottom side (Figure~\ref{fig:rotor}).
During the cool down from ambient temperature to 50\,K, the rotor is held by three retractable linear actuators so that the ring magnet stays above the YBCO pucks by 5\,mm.
The critical temperature of YBCO is $\sim$93 K.
After the cool down procedure, the rotor is released and levitated by the pinning force between the magnet and YBCO.
Rotation is achieved by alternating currents through stator solenoid coils that are magnetically coupled to sprocket neodymium magnets on the edge of the encoder plate. 

We adopt the optical encoder system for the angle reading of the CHWP, as shown in Figure~\ref{fig:encoder}.
Two optical encoder assemblies are installed 180$^\circ$ from each other.
In each assembly, a set of Photo-Diodes (PDs) and IR Light-Emitting Diodes (LEDs) are positioned over and below the rotor's encoder plate, respectively. 
Three of the five PD-LED pairs are for the motor control, while the remaining two pairs are for the rotation angle reading.
The encoder plate has slits to chop the light emitted by the LEDs. 
The photo current signals for the motor control are chopped by the 40 wider slots, while the signals for the angle reading are chopped by the $570 - 1 = 569$ finer slots.
One missing slot is for a reference to mark the rotor's absolute rotation angle. 
The custom driver printed circuit board amplifies the photo current signal and generates electric signals. 
The board flips the direction of the current on the solenoids according to the motor control signal and drives the rotation.
The finer slot signal is fed to the encoder electronics for calculation of the rotation angle of the CHWP. 
We also employ a PID controller to control the HWP speed, and a custom phase compensation unit to optimize the motor drive efficiency.
Further details of the rotation mechanism are illustrated in Ref.~\citenum{yamada2023simons}.

\subsection{Optics}
\begin{figure}
    \centering
    \includegraphics[width = 0.7\textwidth]{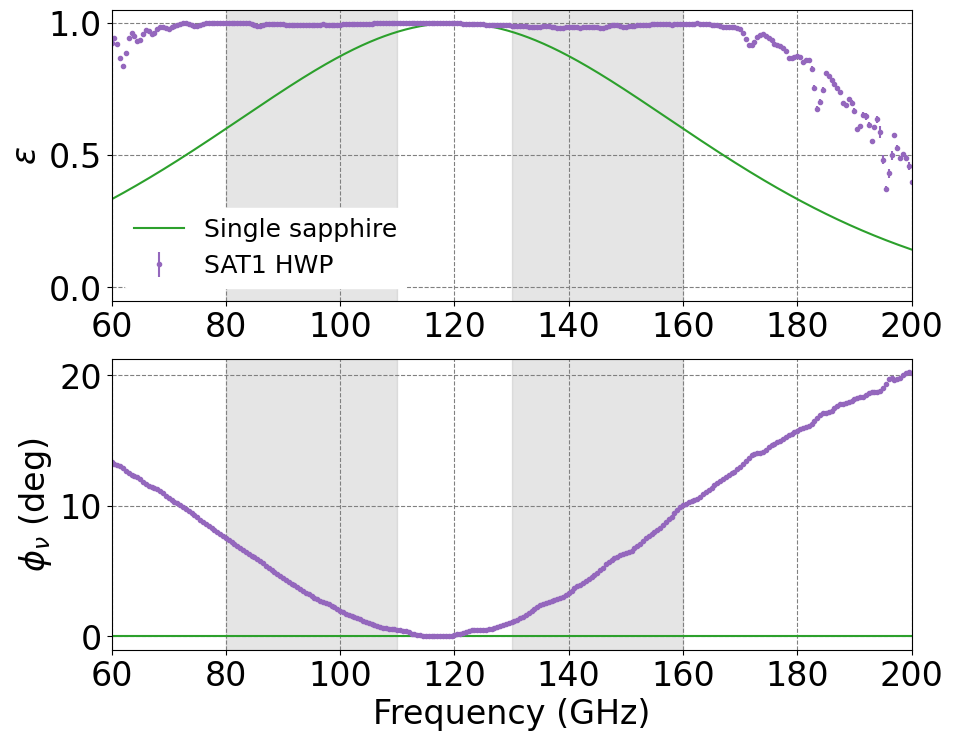}
    \caption{The spectra of the modulation efficiency $\varepsilon$ and phase shift $\phi_\nu$. The green line represents the simulated value of the single sapphire HWP, while the purple points represent the measured value of the SAT1 HWP. The gray bands represent the frequency bands of the SAT1 detectors. }
    \label{fig:ahwp_spectra}
\end{figure}
The HWP of the SATs is made of a-plane sapphire birefringent plates.
Three sapphire plates are stacked in a specific arrangement of the crystal axes to achieve broadband modulation efficiency.\cite{Pancharatnam, Sugiyama2024}
The sapphire stack is sandwiched by the Anti-Reflection (AR) coated alumina plates.\cite{sakaguri2023, Golec_2022}
The AR coating achieves a transmittance of $>90\%$ in the observation bands at cryogenic temperatures, and reduces the instrumental polarization.\cite{Sugiyama2024}

Figure~\ref{fig:ahwp_spectra} shows the frequency dependence of the modulation efficiency $\varepsilon$ and shift in effective crystal axis $\phi_\nu$ of the SAT1 HWP and the theoretical single sapphire plate.
The SAT1 HWP achieves the bandpass-averaged $\varepsilon$ of 99.5\%/98.8\% for the SAT1 (90/150\,GHz band), 97.3\%/99.0\% for the SAT3 (90/150\,GHz band), and 99.8\%/99.7\% for the SAT2 (220/280\,GHz band).\cite{Sugiyama2024}\footnote{These values cannot be directly interpreted as the polarization efficiency of telescopes. It requires considering the spectrum of the input signal, $\phi_\nu$ and transmission of the entire optical system.}
While the crystal axis of birefringence is constant at any frequency for the single sapphire HWP, it depends on frequencies for the 3-layer HWP stack.
The effective shift of the axis is represented by $\phi_\nu$.
The band-averaged $\phi_\nu$ is characterized by the on-site calibration with an expected accuracy of $<0.1^\circ$.\cite{Nakata_2026}.
Further details of the design, fabrication and measurements of the HWP are illustrated in Ref.~\citenum{Sugiyama2024}.

\section{ROTATION ANGLE CALCULATION}
\label{sec:angle_reconstruction}
As previously outlined, the bandpass filtered $d_d(t)$ is multiplied by $2m(-\chi)$ during the demodulation procedure.
In reality, it is impossible to ascertain the true value of $\chi$. 
Instead, the measured HWP rotation angle is employed;
\begin{equation}
    \chi_{meas}(t) = \chi_{true}(t) + \phi(t),
\end{equation}
where $\phi(t)$ is the error in estimation of the HWP angle.
With the angle misestimation, the demodulated signal turns into $(Q(t)+U(t))\exp(i4\phi)$, which means the $2\phi$ rotation of the polarization angle from the true angle. 
This section describes how to estimate the accurate HWP angle~\cite{yamada2025}.

The photo current signal chopped by the slots is fed to the encoder electronics for calculation of the rotation angle of the CHWP. 
The BeagleBone Black microcontroller (BBB) is used for the encoder data acquisition. 
Every time the HWP rotates by one encoder slot, the timestamps of the rising and falling edges of the encoder signal are stored with the 200\,MHz internal clock of the BBB.
The internal clock is further linked to absolute timing information derived from an IRIG-B GPS signal, assigning accurate timestamps to each encoder edge and ensuring synchronization with the detector timestream.


Hereafter, we describe the model for reconstructing the HWP rotation angle from the encoder data.
We assign $j=0, 1$ for the two encoders positioned 180$^\circ$ from one another, as shown in Figure~\ref{fig:sag_model}. 
The $i$-th timestamp measured by encoder $j$ is modeled as
\begin{align}
    t_{i,j} &= t(\chi_{i, j}) \nonumber \\
    &= t\left( \hat{\chi}_i - \phi^{slot}_{i,j} - (-1)^j \phi^{oc}_i \right) \nonumber \\
    &\simeq t\left(\hat{\chi}_i\right) - \frac{dt}{d\chi} \phi^{slot}_{i,j} - (-1)^j \frac{dt}{d\chi} \phi^{oc}_i.
    \label{eq:timestamps_model}
\end{align}
The $\chi_{i, j}$ and $t\left(\hat{\chi}_i\right)$ are the true angle and time when we obtain the timestamp $t_{i,j}$ and incrementally counted apparent angle $\hat{\chi}_i\equiv2\pi i/N$, where $N=1140$ is the total number of rising and falling edges of the encoder slots. 
The index of timestamp $i$ is defined so that $t_{i,j=0} \simeq t_{i,j=1}$, i.e. 0-th timestamp corresponds to the 0-th ($N/2$-th) encoder slot for the encoder~0 (encoder~1).
The $\phi^{slot}_{i,j}$ is an apparent periodic bias of angle due to the deviation of the encoder slot pattern from the ideal equally spaced pattern. 
The following relation of $\phi^{slot}_{i,j}$ holds for all $i$ and $j$:
\begin{align}
\label{eq:slot_period}
    \phi^{slot}_{i,j} = \phi^{slot}_{i+N,j}, \\
    \phi^{slot}_{i,0} = \phi^{slot}_{i+N/2,1}.
    \label{eq:timestamps_model2}
\end{align}
The first equation signifies the periodic nature. 
Under the assumption that the slot pattern in one cycle is consistent during the observation, the same angle bias repeats in every N-th timestamp.
The second equation means that the two encoders see the same slot pattern at a different index by $N/2$.

The $\phi^{oc}_i$ is an apparent angle shift due to the off-centering of the rotor perpendicular to the line connecting two encoder heads. 
This results in an apparent angle offset of opposite signs between the two encoders, as illustrated in Figure~\ref{fig:sag_model}. 
The off-center distance $l^{oc}$ is expressed as
\begin{equation}
\label{eq:offcenter}
    l^{oc} = R_{enc}\tan(\phi^{oc}_i) \simeq R_{enc}\phi_i^{oc},
\end{equation}
where $R_{enc}=$346.25\,mm is the radius of the encoder plate. 

The $d\chi/dt=2\pi f_{HWP}$ stands for the angular speed of the rotor, which is mostly constant at $4\pi$\,rad/s.
The typical size of $|\phi^{slot}_{i, j}|$ is $\lesssim0.1^\circ$, while $\phi^{oc}_i$ varies from $0^\circ$ to $0.7^\circ$ depending on the elevation of the telescope.

\subsection{Slot Pattern Estimation}
First, assuming a smooth rotation and a uniform slot pattern, pseudo
\( t(\hat{\chi}_{i}) \) is estimated by linear interpolation and denoted as 
\( \tilde{t}(\hat{\chi}_{i}) \). 
We take the periodic average of \( t_{i, j} - \tilde{t}(\hat{\chi}_i) \) at each $i$ and obtain the slot pattern degenerately with periodic rotation fluctuation and periodic off-centering.
A more sophisticated method to solve this degeneracy is elaborated in Ref.~\citenum{yamada2025}, and degenerate terms are found to be negligibly small. 

\subsection{HWP Off-center Displacement Estimation}
\label{sec:oc1}
After subtracting the $\phi^{slot}_{i, j}$ term from Equation~\eqref{eq:timestamps_model}, we obtain $t\left(\hat{\chi}_i\right) - (-1)^j \frac{dt}{d\chi} \phi^{oc}_i$ for each encoder.
Since the sign of the off-center term is opposite between the encoder $j=0$ and $1$, the off-center term is canceled by taking average of two encoders.

Conversely, $\phi^{oc}_i$ is derived by subtracting the data of two encoders from each other.
We can also estimate the off-center distance $l^{oc}$ with derived $\phi^{oc}_i$ and Equation~\eqref{eq:offcenter}.
This is utilized to estimate the rotor position for CHWP operation safety.
We can also evaluate the vibration and the spring constant of the SMB from the off-center measurement\cite{yamada2023simons}. 

\section{ON-SITE OPERATION AND PERFORMANCE}
\subsection{Observation History and Uptime}

\begin{figure}
    \centering
    \includegraphics[width = 1\textwidth]{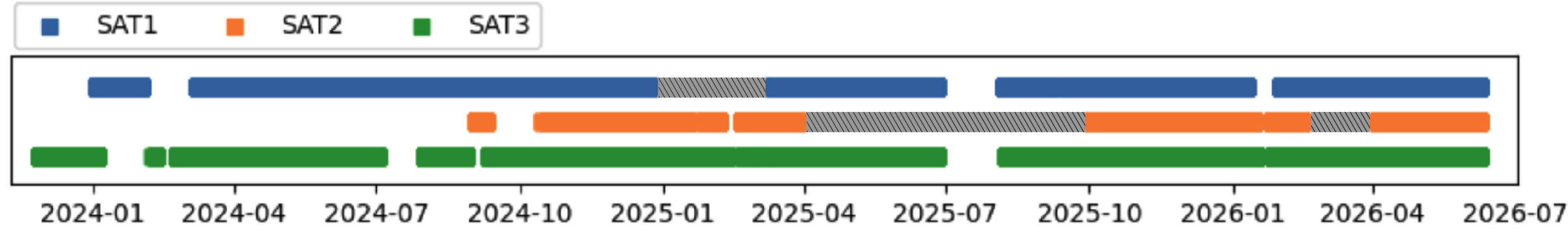}
    \caption{The duration for which the CHWPs of each SAT are in operation. The durations marked with slashes indicate the telescope retrofits. The other periods during which the CHWP was not running correspond to the periods during which the SAT was not observing due to maintenance or shutdown.}
    \label{fig:operation}
\end{figure}

Figure~\ref{fig:operation} illustrates the observational periods for each SAT, during which the CHWP is running at 2\,Hz.
The three CHWPs have been continuously in operation since their deployment, aside from suspensions for the retrofit and power shut down. 
The robust rotation control allows the months of continuous operation.

\subsection{CHWP angle noise}
\begin{figure}
    \centering
    \includegraphics[width = 0.8\textwidth]{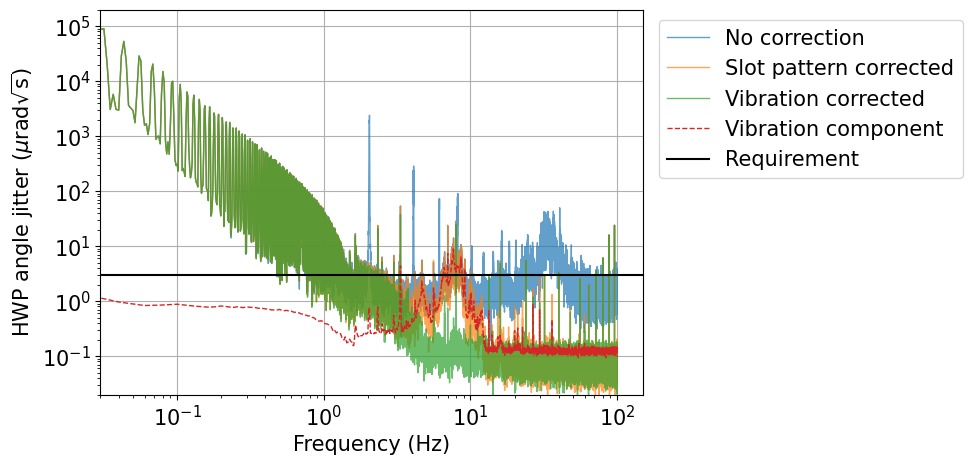}
    \caption{The ASD of the HWP angle jitter from a representative 1 hour observation, when the telescope is scanning the sky at 0.5\,deg/s and rotating the HWP at 2\,Hz. The blue, orange and green lines represent the angle jitter before corrections, after the slot pattern correction, and after both slot pattern and off-center (i.e. vibration) corrections, respectively. The red dashed line shows the angle jitter induced by the vibration. The corrections subtract the angle jitter successfully and suppress the angle uncertainty to less than 3\,$\mu$rad$\sqrt{\text{s}}$ in the signal band around the modulation frequency.}
    \label{fig:angle_jitter}
\end{figure}
Here, we evaluate the rotation stability and the angle measurement accuracy of the CHWP.
The upper limit of 3\,$\mu$rad$\sqrt{\text{s}}$ was established as the requirement for the encoder white noise level \cite{yamada2023simons}. 
Figure~\ref{fig:angle_jitter} represents the ASD of the CHWP angle jitter for reconstructed angles before and after corrections, which are measured when the SAT is scanning the sky at 0.5\,deg/s.
The noise ASD of the uncorrected angle has peaks at 2\,Hz and its harmonics.
They are caused by the periodic bias of the rotation angle, and are cleaned by the slot pattern subtraction.
The larger white noise and the broad peak around 40\,Hz are induced by the slot pattern non-uniformity, which are also subtracted by the slot pattern subtraction.
The broad peak around 10\,Hz is due to the scan-induced vibrations of the rotor, which is subtracted by the off-center correction.
The motions of the telescope excite vibrations of the rotor at characteristic vibration frequencies of the superconducting magnetic bearing. 
This is an apparent change of rotation angle, and the flat white noise profile after its subtraction validates the correction process.
With corrections, the typical noise level is reduced to 0.16\,$\mu$rad$\sqrt{\text{s}}$.
The requirement of $<$3\,$\mu$rad$\sqrt{\text{s}}$ is achieved in more than 99.9\% of observations.

\subsection{Optical noise induced by CHWP angle jitter}
\label{sec:HWPSS}
Equation~\eqref{eq:modulation} represents the signal modulation of the ideal HWP.
However, the real HWP generates HWP Synchronous Signal (HWPSS) in addition to the ideal response due to its optical and mechanical imperfections.
The HWPSS is expressed as
\begin{equation}
\label{eq:hwpss}
    \mathcal{P}_\mathrm{HWPSS}=\sum_{n=1}\mathrm{Re}[A_n(t)e^{-in\chi}],
\end{equation}
where the $A_n$ is the $n$-th amplitude of the HWP harmonic.
The HWPSS appears in the modulated signal, as shown in Figure~\ref{fig:chwp2}.
The peaks at 2, 4, 6, 8~Hz are the $n=1, 2, 3, 4$ component, respectively.
The $n=4$ component needs to be carefully estimated because it stands at the same frequency as the modulated polarization signal.

The HWPSS produces largely common noise on the detectors in conjunction with the HWP rotation angle jitter.
Assuming that the HWP rotation angle has an uncertainty of $\Delta\chi(t)$, the $n=4$ component of Equation~\eqref{eq:hwpss} becomes
\begin{align}
    \mathcal{P}_\mathrm{4f}
    &=\mathrm{Re}[A_4~e^{-i4\left(\chi+\Delta\chi(t)\right)}]\nonumber \\
    &=A_4\cos{\left(4(\chi+\Delta\chi(t))\right)}\nonumber \\
    &\simeq A_4\cos{4\chi} - 4\Delta\chi(t)A_4\sin{(4\chi}),
\end{align}
where the second term in the third line appears as noise.\footnote{In reality, all HWPSS components induce the noise. Here we evaluate only $n=4$ as it is the first order component.}
The Root Mean Square (RMS) of this noise is $2\sqrt{2}A_4\sigma_\chi$, where the $\sigma_\chi$ is the RMS of the HWP angle jitter.
This noise is largely correlated between detectors and does not average down with the detector count. 
Therefore, it must be suppressed to be less than the Noise Equivalent Power of the detector array, $\mathrm{NEP^{arr}}$.

\begin{figure}
    \centering
    \includegraphics[width = 1\textwidth]{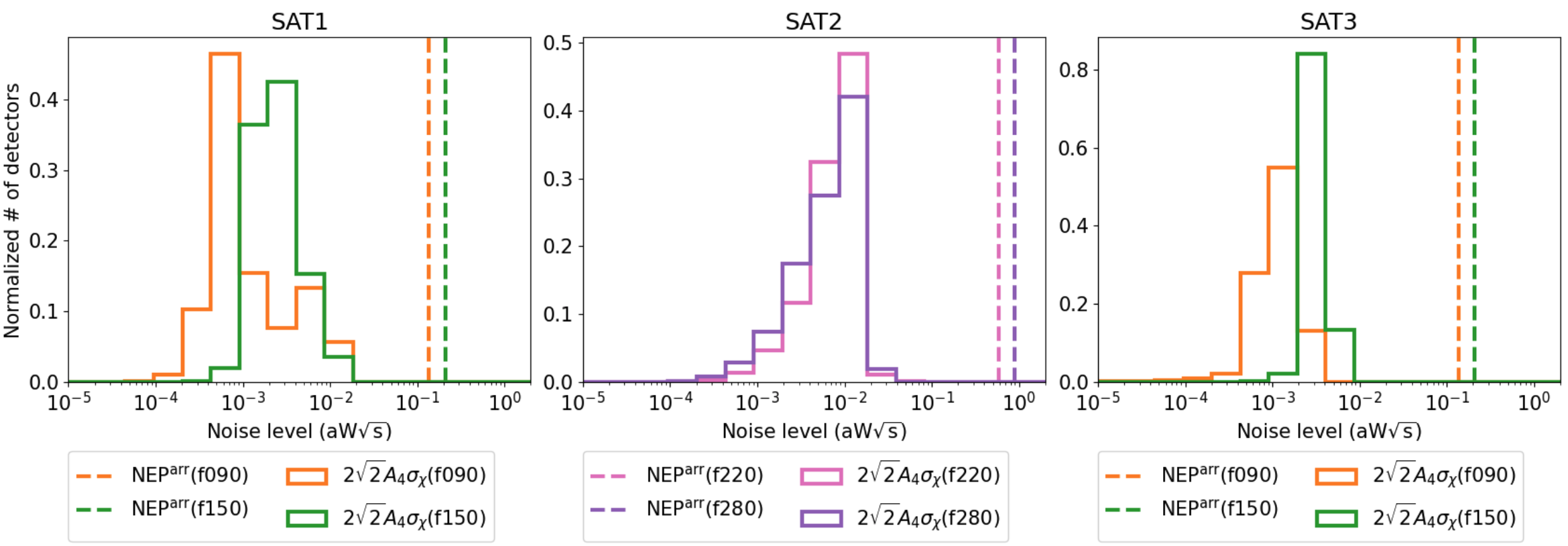}
    \caption{The histogram shows the expected white noise level induced by HWPSS, given the RMS of the HWP angle jitter of $\sigma_\chi=0.16\,\mu\mathrm{rad}\sqrt{\mathrm{s}}$. The histogram is scaled by the total number of available detectors. The vertical lines show the expected $\mathrm{NEP^{arr}}$ of the SATs, assuming that all the 6048 detectors are in operation. The optical noise induced by the CHWP is smaller than $\mathrm{NEP^{arr}}$ by an order of magnitude at maximum.}
    \label{fig:hwpss2}
\end{figure}

Figure~\ref{fig:hwpss2} shows the expected noise RMS induced by the CHWP. 
The vertical lines represent the $\mathrm{NEP^{arr}}$ of each frequency band, assuming that 6048 detectors are in operation~\cite{Dutcher2024, 11397458}.
The measured $A_4$ depends on the detectors, and $2\sqrt{2}A_4\sigma_\chi$ on each detector is shown in the histogram.
The maximum $2\sqrt{2}A_4\sigma_\chi$ is smaller than $\mathrm{NEP^{arr}}$ by an order of magnitude at every frequency band.
Thanks to the control of the HWPSS and CHWP angle jitter, the noise RMS induced by the CHWP is quite negligible. 

\subsection{HWP Off-center Measurement}
\label{sec:offcenter}
The angular offset caused by off-center displacement of the rotor $\phi^{oc}$ is important as it primarily depends on the elevation of the telescope and introduces an elevation dependent bias of the measured polarization angle. 
This has to be corrected by the off-center measurement described in Section~\ref{sec:oc1} with an accuracy of  $<0.1^\circ$, which is posed by the SATs' calibration requirement of polarization angle \cite{10.1117/12.2313832}.

\begin{figure}
    \centering
    \includegraphics[width = 0.45\textwidth]{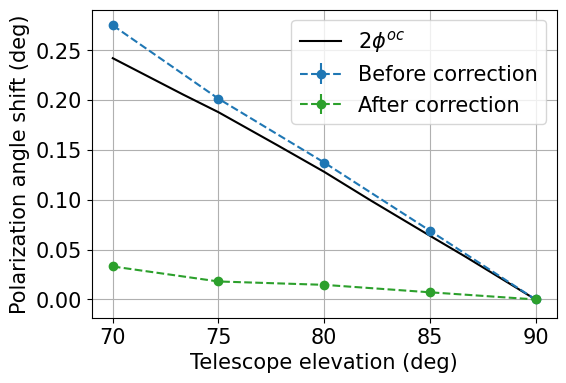}
    \caption[HWP off-center measurement]{The polarization from a SWG is measured at telescope elevations from 70 to 90 degrees. The measured polarization angles with or without correction of $\phi^{oc}$ is plotted as a function of the elevation. Without correction, the reconstructed polarization angle rotates depending on elevation. The reconstructed polarization angle with off-center correction is almost constant in the elevation change.}
    \label{fig:sag}
\end{figure}

To optically verify the correction of $\phi^{oc}$, we performed following measurements. 
We placed the Sparse Wire Grid (SWG), the linear polarization source for the calibrations~\cite{Murata2023, Nakata_2026}, with a fixed angle on top of the window to input a constant linear polarization.
Then we changed the elevation of the telescope from 90$^\circ$ to 70$^\circ$ by the 5$^\circ$ step to induce the change in $\phi^{oc}$. 
The solid black line of Figure~\ref{fig:sag} shows the expected shift of polarization angle due to $\phi^{oc}$. 
The rotor is off-centered by approximately 0.73\,mm from elevation 90$^\circ$ to 70$^\circ$, which corresponds to the polarization angle shift of $2\phi^{oc}=0.24^\circ$. 
The blue and green lines in Figure~\ref{fig:sag} represent the reconstructed polarization angle with respect to the telescope elevation, with and without correction of $\phi^{oc}$.
As expected, the measured polarization angle shifts depending on elevation without correction. 
After correction, the measured polarization angle is almost constant. 
The residual polarization angle shift of $<0.05^\circ$ may be attributable to the increase of detector time constants and the sag of wire grid at lower elevations.\footnote{As the telescope elevation goes down, it sees the thicker atmosphere, therefore gets larger loading on its detectors and results in the larger time constant. The time constant affects the demodulated polarization angle. The wire angle of SWG also slightly changes with the telescope elevation.}
The residuals are not necessarily problematic because SATs perform the detector calibration before they operate the constant elevation scan.

\section{CONCLUSION}
The SATs of the SO experiment explore the degree-scale CMB $B$-mode, which may originate from primordial gravitational waves.
The CHWPs are installed in the SATs to suppress the $1/f$ noise, which contaminates the degree scale $B$-mode signal.
They reduce the effect of low-frequency noise in the intensity measurement by orders of magnitude in the polarization measurement.
The accurate estimation of the CHWP rotation angle is essential to minimize the systematic uncertainty on the polarimetry.
We have developed optical angle encoders and a precise angle reconstruction model.
The performance of the angle reconstruction is evaluated for the SATs in operation at the observation site.
The jitter of the reconstructed rotation angle is evaluated to be 0.16\,$\mu\mathrm{rad\sqrt{s}}$, which meets the requirement.
The effective white noise on the detectors induced by this angle jitter is confirmed to be negligible at all the SATs.
We have also optically validated the angle offset correction.
The contactless, robust and accurate rotation control system enables the stable long-term operation with minimal systematics.

\acknowledgments 
This work was supported by MEXT KAKENHI Grant
Numbers JP19H00674, JP23H00105, JP23H01202, \\
JP22H04913, and by JSPS KAKENHI Grant Numbers JP24K23938, JP24KJ0663, JP25KJ0840, and by the JSPS Core-to-Core Program JPJSCCA20200003. This
work was supported by World Premier International
Research Center Initiative (WPI). This work was supported in part by a grant from the Simons Foundation
(Award $\#$457687, B.K.). Work at Princeton was supported by the David Wilkinson Research Fund. Work at
LBNL was supported in part by the U.S. Department of
Energy, Office of Science,Office of High Energy Physics,
under contract No. DE-AC02-05CH11231. The work
reported on in this paper was performed using Princeton University’s Research Computing resources and resources of the National Energy Research Scientific Computing Center (NERSC). J.S. acknowledges the support
from the International Graduate Program for Excellence
in Earth-Space Science (IGPEES) and the JSR Fellowship, the University of Tokyo. D.S. acknowledges the
support from FoPM, WINGS Program and the JSR Fellowship, the University of Tokyo. We thank Jansen Ball
for improving the motor drive electronics.  

\bibliography{reference} 
\bibliographystyle{spiebib} 

\end{document}